\documentclass[aip, jap, reprint]{revtex4-1}

\usepackage{hyperref}
\usepackage{graphicx}
\usepackage{amsmath}
\usepackage{wasysym}
\usepackage{amsfonts}
\usepackage{color}
\usepackage{verbatim}

\graphicspath{{./figures/}}

\begin{document}
\title{Compact femtosecond electron diffractometer with 100 keV electron bunches approaching the single-electron pulse duration limit}
\author{L. Waldecker} 
\email{waldecker@fhi-berlin.mpg.de}
\affiliation{Fritz-Haber-Institut der Max-Planck-Gesellschaft, Berlin, Germany.}
\author{R. Bertoni}
\affiliation{Fritz-Haber-Institut der Max-Planck-Gesellschaft, Berlin, Germany.}
\author{R. Ernstorfer}
\email{ernstorfer@fhi-berlin.mpg.de}
\affiliation{Fritz-Haber-Institut der Max-Planck-Gesellschaft, Berlin, Germany.}

\begin{abstract}

We present the design and implementation of a highly compact femtosecond electron diffractometer working at electron energies up to 100 keV. We use a multi-body particle tracing code to simulate electron bunch propagation through the setup and to calculate pulse durations at the sample position. Our simulations show that electron bunches containing few thousands of electrons per bunch are only weakly broadened by space-charge effects and their pulse duration is thus close to the one of a single-electron wavepacket. With our compact setup we can create electron bunches containing up to 5000 electrons with a pulse duration below 100 femtoseconds on the sample. We use the diffractometer to track the energy transfer from photoexcited electrons to the lattice in a thin film of titanium. This process takes place on the timescale of few-hundred femtoseconds and a fully equilibrated state is reached within one picosecond. 

\end{abstract}

\maketitle
\date{\today}

\section{Introduction}
\label{sec:intro}

The ultrafast dynamics of matter after photoexcitation are governed by the interactions of electrons, nuclei and spins. High excitation levels can induce phase transitions or drive exotic non-thermal processes, whereas weak excitations can be regarded as perturbations of the equilibrium state. Time-resolved experiments using the pump-probe scheme allow studying these photo-induced processes on their fundamental time-scale. Optical- and photoemission experiments indirectly or directly measure electronic properties\cite{Book:Bovensiepen}, whereas the evolution of the atomic structure can be probed by short x-ray or electron pulses. These techniques directly measure atomic order and combine femtosecond temporal- with subatomic spatial resolution. \\
While high-flux x-ray sources require large scale facilities, such as free-electron lasers or synchrotons, the development of femtosecond electron diffraction (FED) experiments has led to very compact table-top setups capable of visualizing phenomena like lattice melting \cite{Siwick2003, Ernstorfer2009}, melting of a charge density wave \cite{Eichberger2010}, structural changes related to a metal-insulator transition \cite{Morrison2014} and breathing motion of thin films \cite{Park2005}.  \\
Despite a lot of ongoing development, time-resolution of electron diffraction experiments has so far been limited by electron bunch durations on the sample position, which typically ranged from few picoseconds in early setups\cite{Williamson1984, Dantus1994} to few-hundreds of femtoseconds in some recent designs \cite{Kassier2010, Hebeisen2008}. Whereas calculations have shown that short pulse durations are easily achievable for single-electron pulses \cite{Aidelsburger2010}, it remains a challenge to achieve short bunch durations with a higher number of electrons.  The repulsive coulomb-interaction between electrons leads to their spatial separation during propagation, what translates to temporal broadening of the electron bunch \cite{Siwick2002}. Schemes for reversing the typical linear momentum spread and compressing electron pulses have been proposed \cite{Fill2006,VanOudheusden2007,Kassier2009} and a scheme using a radio frequency cavity has been implemented recently \cite{Chatelain2012, Gao2012, Mancini2012}. Up to now, time resolution in those experiments is limited due to jitter in the synchronization of RF-cavity and the laser pump pulses\cite{Chatelain2012, Gao2012}. \\
In this paper we present a setup for ultrafast electron diffraction that aims at minimizing electron propagation distances and therefore circumvents the need of a compression scheme. We present simulations that suggest that, with the geometry at use, we are able to get electron pulses of around 100 fs at the sample position for $10^3$ - $10^4$ electrons per pulse. To demonstrate the capabilities of our setup, we measure the time-constant of the lattice heating of titanium following photo-excitation. We show that the energy transfer from electrons to the lattice takes place on the timescale of 350 femtoseconds and a thermalized state is reached within one picosecond. \\

\section{Femtosecond electron diffraction setup}
\label{sec:setup}

The ultrafast electron diffractometer is based on a pump-probe measurement scheme. A femtosecond laser pulse is split into two parts. The pump pulse excites the thin film sample under study and its structure is probed by a transmitting electron pulse which diffracts off it. Electron probe pulses are created by photoemitting electrons from a thin metal film by the same laser and are accelerated in a static electric field. Pump and probe pulses are therefore intrinsically optically synchronized. \\
\begin{figure}[ht!]
\begin{center}
\includegraphics[width=1.0\columnwidth]{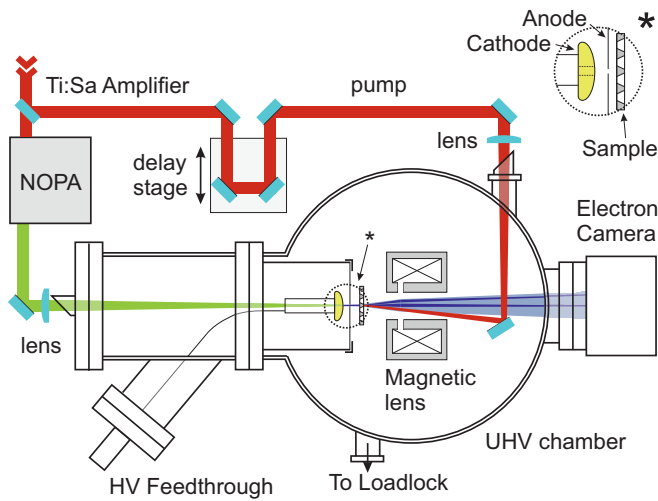}
\caption{Schematic of the experimental setup for femtosecond electron diffraction (detailed explanation in text).}
\label{fig:setup}
\end{center}
\end{figure}
Figure~\ref{fig:setup} shows a schematic of the table-top electron diffractometer. The output of a Ti:sapphire amplified laser system (1~mJ / 30~fs) is split into a pump- and a probe arm. The probe-arm drives a non-collinear optical parametric amplifier (NOPA)\cite{Wilhelm1997} that produces pulses of 30~fs pulse duration tunable in the visible spectral region. Its output is used to photo-emit electrons from a photocathode in a two-photon process. Subsequently, the created electron pulse is accelerated in a static electric field created by the voltage difference between the photocathode and the grounded anode. The anode-cathode distance being 8 mm, the field strength at the maximum voltage of 100 kV reaches 12.5 MV/m. With the field for vacuum break-through of a perfectly homogeneous surface being on the order of 20 MV/m \cite{book:Latham}, care had to be taken in the design of the cathode/anode assembly as well as in the quality of the surfaces of all parts of the electron gun to avoid field emission and arcing. The shape of the cathode is an approximation to the shape of the uniform field electrode described by Bruce \cite{Bruce1946}. A flat, 5 mm diameter sapphire plate is fitted in a hole in the center of the cathode and the assembly is then coated with a 20-30 nm thick gold film. The electrons are emitted from this thin film by illuminating the photocathode from the back side with the output of the NOPA. Since transverse and longitudinal coherence in the electron bunch increase for lower excess energy in the emission process \cite{Michalik2008}, the  central wavelength of the NOPA is chosen to be just above the work function of the gold cathode. The Schottky-effect lowers the effective work function by roughly $\Delta W \approx 130$ meV at 100 kV and it is found to be approximately 4.4 eV.  The anode is built out of a polished 4 inch Si-wafer with a 150 $\mathrm{\mu m}$ diameter hole in its center, through which the accelerated electrons leave the gun and propagate towards the sample. \\
The diffraction experiments are performed in transmission geometry. Therefore, thin samples are required with typical thicknesses that range from few atomic layers to less than 100~nm. We use standard TEM-grids as well as specially designed grids for large area samples allowing for single-shot studies of irreversible processes\cite{Waldecker2014}. The samples can be translated in three dimensions with a motorized xyz-translation stage, and tip and tilt angles can be controlled by two additional motors. The distance of the sample to the gun can be varied, with the minimum distance of the sample to the anode being 2 mm. This makes the entire propagation distance of the electron bunch from cathode to the sample as short as 10 mm. \\
In contrast to early designs \cite{Dwyer2006}, the electron pulse diffracts off the sample without going through a focusing element beforehand. We employ an FED layout introduced recently\cite{Gerbig2012} where the diffracted electrons are focused onto the detector with a magnetic lens only placed behind the sample. This design significantly decreases the electron bunch propagation distance and avoids temporal distortions induced by the magnetic lens itself\cite{Weninger2012}. The position of the lens is controlled by a three-axis manipulator, with which we place the lens approximately 5 cm behind the sample.  The lower limit of the transverse coherence length of the electron bunches on the sample is estimated to be 4 nm by comparison of the peak width to peak separation\cite{Kirchner2013} of a standard diffraction sample (32 nm Al, Plano GmbH).  \\
This configuration additionally enables imaging the sample on the detector, which allows for easy sample positioning and characterization. Diffraction patterns and magnified real-space images are detected with a commercial electron camera  (TVIPS TemCam F416), capable of every-electron detection. \\
The pump pulses are sent over a delay stage that changes the relative arrival time of pump and probe, and is then focused onto the sample. In our compact design, the pump pulses hit the sample from the opposite side as the probe. As samples are very thin, the excitation profile in the sample is homogeneous for most materials and differences in arrival time due to counter-propagating pulses can be neglected. The angle of the pump pulses with respect to electron propagation axis is kept small (below 5 degrees) to reduce degradation of the temporal resolution to below 30 fs. The photocathode potential is stable to $10^{-5}$ of the maximum voltage (power supply: Heinzinger PNChp 100000), and jitter in arrival time of electron pulses is approximately one femtosecond.  \\

\section{Simulation of electron pulse propagation}
\label{sec:epulse}

\begin{figure*}[bt]
\begin{center}
\includegraphics[width=0.95\textwidth]{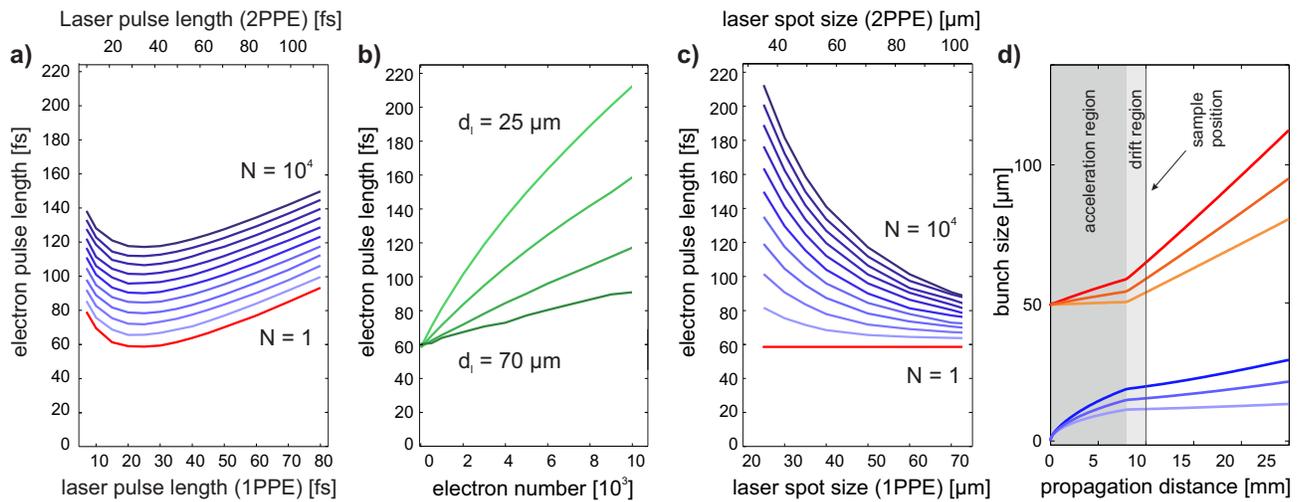}
\caption{Results of the GPT simulation. a) Electron bunch duration as function of  laser pulse duration and number of electrons per pulse (red: one electron, blue: $10^3$ to $10^4$ electrons in steps of $10^3$, light to dark). b) Electron bunch duration on the sample as function of electron number for laser spotsizes of $d_l=25$ (light), 35, 50 and 70 $\mu$m  (dark). c) Electron bunch duration as a function of laser spotsize and different numbers of electrons per pulse (colors as in a)) d) transversal (red dashed) and longitudinal (blue) FWHM bunch sizes for $10^3$, $5*10^3$ and $10^4$ electrons (light to dark). }
\label{fig:sim}
\end{center}
\end{figure*}

The interaction of electrons via Coulomb forces makes the description of the propagation of multi-electron pulses a non-trivial problem. Here, we use a fully relativistic multi-body particle tracing code (general particle tracer (GPT), Pulsar Physics) that includes full particle-interactions to simulate the propagation of a bunch of charged macroparticles through our setup and discuss how experimentally accessible parameters influence the pulse duration at the sample position. We give the electron pulse duration as approximate FWHM, calculated as 2.35 times the standard deviation of the arrival times of electrons at the sample. We find that using 1000 macro-particles is sufficient for converged simulations of the propagation of pulses containing up to $10^4$ electrons. The macro-particles are created at different simulation times around zero with a Gaussian temporal distribution of FWHM $\tau_l$. They enter a static electric field of 12.5 MV/m, created by two flat surfaces, the anode containing a hole of 150 $\mu$m diameter. The electron bunch is assumed to have Gaussian distribution of particle density in both transverse directions, with an initial width of $d_l$.  Each particle is assigned an energy and momentum at creation. The momentas directions are distributed uniformly on a half-sphere in forward propagation-direction. Energies are assumed to follow a Gaussian distribution as expected for electrons photoemitted by a laserpulse with a Gaussian spectrum. Its width is calculated from the laser pulse-length $\tau_l$ by assuming the pulse to be Fourier-limited. To account for inhomogeneities of the photocathode, we convolute this energy with a Gaussian distribution of 100~meV width and obtain $\Delta E_{\mathrm{initial}} = ({\left(0.44*h/\tau_l\right)^2 + 0.1^2} )^{1/2}$. The center of the distribution is one sigma above zero. \\

The duration of single-electron pulses depends on emission bandwidth $\Delta E_{\mathrm{initial}}$, laser pulse length and the electric field in the acceleration region. For a given field strength, there is a tradeoff between bandwidth and laser pulse duration and the shortest electron pulses are achieved for a specific time-bandwidth combination \cite{Aidelsburger2010}. Pulses containing more than one electron acquire an additional energy spread during their propagation due to Coulomb interactions between the particles. The dependence of electron pulse length at the sample position, i.e.~at a distance of 10~mm from the cathode, on electron number and laser pulse duration, calculated with the GPT code, is shown in figure \ref{fig:sim} a). The upper axis has been added to show the effect of photoemission in a two-photon process (two photon photoemission, 2PPE) as used in our laboratory. The effective pulse length is lowered by a factor of $\sqrt{2}$ due to the nonlinearity of the process, whereas the bandwidth is increased by the same factor, what results only in a shift of the abscissa. The laser spot size $d_l$ was chosen to be 50 $\mu$m (1PPE), corresponding to 71 $\mu$m (2PPE), close to the typical experimental value.  The considered range of one to $10^4$ electrons corresponds to the experimentally accessible range with our setup. \\
Simulations without space charge (red dashed curve) qualitatively reproduce the expected shape of the analytical expression\cite{Aidelsburger2010}. Pulses containing more than one electron follow similar curves (blue lines), with pulse durations shifted to higher values for higher electron numbers. For our experimental parameters, the overall space-charge induced broadening of the pulselength is 10\% for a pulse containing $10^3$ electrons and 100\% for $10^4$ electrons. \\
Figures \ref{fig:sim}~b) and \ref{fig:sim}~c) show the dependence of bunch duration at the sample position on electron number per bunch and laser spot size calculated with our N-body particle tracing simulation for the optimal laser pulse length ($\tau_{l, \mathrm{1PPE}} = 25$~fs). A sublinear dependence of bunch duration with electron number is found. Increasing the spot size on the photocathode results in a reduction of electron bunch duration on the sample. Even though increasing the bunch diameter is advantageous for having short bunch durations, there are practical limits in increasing the electron beam size, e.g. increasing temporal mismatch of non-collinear pump and probe pulses and finite sample sizes. \\
Several analytical models describing electron bunch propagation have been proposed in recent years\cite{Siwick2002, Qian2003}. We find a qualitative agreement of our simulations with the predictions of these models for small electron numbers per pulse, but deviations for higher electron numbers. A quantitative comparison is difficult even for low electron numbers, because of the geometrical assumption of a cylindrical bunch shape in the analytical models. \\
The models assume that transversal distances between electrons directly after emission are typically much bigger than distances in longitudinal (propagation) direction ($r_b \gg l$). Figure \ref{fig:sim} d) shows longitudinal and transversal bunch diameters during the propagation through the setup. Both change dynamically, opposed to the assumptions of the models. Especially, passage of the electron bunch through the anode changes the divergence of the bunch, because of the hole acting as an Einzel-lens. For higher electron numbers, longitudinal and transversal bunch sizes become comparable in magnitude in conflict with the assumption $r_b \gg l$ of these  models. This illustrates the importance of using multi-body simulations for the accurate calculation of the propagation of femtosecond electron bunches.

\section{Measurement of the ultrafast lattice heating of titanium}

\begin{figure}[ht!]
\begin{center}
\includegraphics[width=1.0\columnwidth]{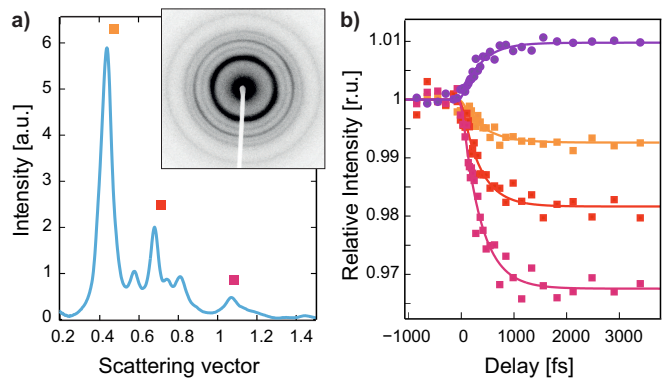}
\caption{(a) Radial average and raw diffraction image of titanium. In the radial average, a background has been subtracted. (b) Relative peakheight of three selected peaks (squares) and of the integrated background (circles) with respect to negative time delays.}
\label{fig:Tidecay}
\end{center}
\end{figure}

We demonstrate the capabilities of the FED setup by measuring the time-constant of the energy transfer from excited electrons to the lattice in titanium. The sample used is a 30 nm thick film of polycristalline titanium that was grown on a NaCl single crystal and then transferred onto a standard TEM-grid. We used the fundamental of the amplified laser system at an incoming fluence of 4 $\mathrm{mJ/cm^2}$ to excite the sample. The intensity of the NOPA was adjusted to photo-emit about 1000 electrons per pulse. \\
In figure \ref{fig:Tidecay} a) the radial average of titanium is shown, obtained by angularly integrating the two-dimensional diffraction image shown in the inlet. The image is integrated over 2500 pulses and a background has been fitted and subtracted from the radial average.   \\

We fit the peaks with pseudo-Voigt line-profiles to determine their position and intensity.  The resulting relative change of intensity of three selected Bragg-peaks as well as the relative change in the integrated background intensity is plotted as a function of pump-probe delay in figure~\ref{fig:Tidecay}~b). The data is averaged over 30 scans, each integrated for 5 s per delay point at a laser repetition rate of 500 Hz. The solid lines are single-exponential fits, convoluted with a 100 fs FWHM Gaussian function to account for our expected time-resolution, based on the experimental parameters and the simulations of the electron pulse duration. Immediately after photoexcitation, energy is transferred from electrons to the lattice, which tend towards a new thermal equilibrium at an elevated new temperature $T_{\mathrm{final}}$. At this new temperature, the Debye-Waller factor and therefore the relative height of the peaks is decreased due to a bigger mean square displacement $\left \langle u^2 \right \rangle$ of the atoms around their equilibrium positions. At the same time the background is increased due to inelasticly scattered electrons. The fits yield a timeconstant $\tau = 360\pm 31$ fs for the decrease of the peaks, and a simmilar timeconstant of  $\tau = 315\pm 45$ fs for the increase of the integrated background. The new equilibrium is reached within one picosecond, after which the peakheights stays constant in our measurement range of several tens of picoseconds but recover on the timescale of the 500 Hz repetition rate of our laser.    \\

\section{conclusion}
We have implemented a table-top femtosecond electron diffraction setup that follows the approach of miniaturizing propagation distances of electron probe pulses to minimize broadening caused by vacuum dispersion and Coulomb-repulsion. The diffraction setup works at electron energies up to 100 keV. We have performed multi-particle simulations which suggest that electron pulse durations on the sample can be as short as 60 fs for single-electron pulses. Increasing the number of electrons per pulse to $10^3$ does broaden pulses by only 10\%  and pulses with up to $5*10^3$ electrons can have pulse durations below 100 fs. Since there is no synchronization needed, our time resolution is close to the electron bunch duration. To our knowledge, this is the first setup with multi-electron bunches allowing for time-resolution less than 100 fs. We have demonstrated the capabilities of our setup by measuring the ultrafast dynamics of the lattice heating of titanium, caused by photoexcitation with a femtosecond laser pulse and the subsequent interaction of electrons and phonons.

\section{acknowledgments}
Titanium samples were produced by Dhriti Ghosh and Valerio Pruneri at the ICFO Barcelona. The authors acknowledge discussions with Germ\'{a}n Sciaini. This research was funded by the Max Planck Society. L. W. acknowledges support by the Leibniz graduate school 'Dynamics in New Light'. R. B. thanks Alexander von Humboldt foundation for financial support.

\bibliography{Setup_literatur}

\begin{thebibliography}{28}%
\makeatletter
\providecommand \@ifxundefined [1]{%
 \@ifx{#1\undefined}
}%
\providecommand \@ifnum [1]{%
 \ifnum #1\expandafter \@firstoftwo
 \else \expandafter \@secondoftwo
 \fi
}%
\providecommand \@ifx [1]{%
 \ifx #1\expandafter \@firstoftwo
 \else \expandafter \@secondoftwo
 \fi
}%
\providecommand \natexlab [1]{#1}%
\providecommand \enquote  [1]{``#1''}%
\providecommand \bibnamefont  [1]{#1}%
\providecommand \bibfnamefont [1]{#1}%
\providecommand \citenamefont [1]{#1}%
\providecommand \href@noop [0]{\@secondoftwo}%
\providecommand \href [0]{\begingroup \@sanitize@url \@href}%
\providecommand \@href[1]{\@@startlink{#1}\@@href}%
\providecommand \@@href[1]{\endgroup#1\@@endlink}%
\providecommand \@sanitize@url [0]{\catcode `\\12\catcode `\$12\catcode
  `\&12\catcode `\#12\catcode `\^12\catcode `\_12\catcode `\%12\relax}%
\providecommand \@@startlink[1]{}%
\providecommand \@@endlink[0]{}%
\providecommand \url  [0]{\begingroup\@sanitize@url \@url }%
\providecommand \@url [1]{\endgroup\@href {#1}{\urlprefix }}%
\providecommand \urlprefix  [0]{URL }%
\providecommand \Eprint [0]{\href }%
\providecommand \doibase [0]{http://dx.doi.org/}%
\providecommand \selectlanguage [0]{\@gobble}%
\providecommand \bibinfo  [0]{\@secondoftwo}%
\providecommand \bibfield  [0]{\@secondoftwo}%
\providecommand \translation [1]{[#1]}%
\providecommand \BibitemOpen [0]{}%
\providecommand \bibitemStop [0]{}%
\providecommand \bibitemNoStop [0]{.\EOS\space}%
\providecommand \EOS [0]{\spacefactor3000\relax}%
\providecommand \BibitemShut  [1]{\csname bibitem#1\endcsname}%
\let\auto@bib@innerbib\@empty
\bibitem [{\citenamefont {Bovensiepen}, \citenamefont {Petek},\ and\
  \citenamefont {Wolf}(2012)}]{Book:Bovensiepen}%
  \BibitemOpen
  \bibinfo {editor} {\bibfnamefont {U.}~\bibnamefont {Bovensiepen}}, \bibinfo
  {editor} {\bibfnamefont {H.}~\bibnamefont {Petek}}, \ and\ \bibinfo {editor}
  {\bibfnamefont {M.}~\bibnamefont {Wolf}},\ eds.,\ \href@noop {} {\emph
  {\bibinfo {title} {Dynamics at Solid State Surfaces and Interfaces Volume 2:
  Fundamentals}}}\ (\bibinfo  {publisher} {Wiley-VCH},\ \bibinfo {year}
  {2012})\BibitemShut {NoStop}%
\bibitem [{\citenamefont {Siwick}\ \emph {et~al.}(2003)\citenamefont {Siwick},
  \citenamefont {Dwyer}, \citenamefont {Jordan},\ and\ \citenamefont
  {Miller}}]{Siwick2003}%
  \BibitemOpen
  \bibfield  {author} {\bibinfo {author} {\bibfnamefont {B.~J.}\ \bibnamefont
  {Siwick}}, \bibinfo {author} {\bibfnamefont {J.~R.}\ \bibnamefont {Dwyer}},
  \bibinfo {author} {\bibfnamefont {R.~E.}\ \bibnamefont {Jordan}}, \ and\
  \bibinfo {author} {\bibfnamefont {R.~J.~D.}\ \bibnamefont {Miller}},\ }\href
  {\doibase 10.1126/science.1090052} {\bibfield  {journal} {\bibinfo  {journal}
  {Science}\ }\textbf {\bibinfo {volume} {302}},\ \bibinfo {pages} {1382}
  (\bibinfo {year} {2003})}\BibitemShut {NoStop}%
\bibitem [{\citenamefont {Ernstorfer}\ \emph {et~al.}(2009)\citenamefont
  {Ernstorfer}, \citenamefont {Harb}, \citenamefont {Hebeisen}, \citenamefont
  {Sciaini}, \citenamefont {Dartigalongue},\ and\ \citenamefont
  {Miller}}]{Ernstorfer2009}%
  \BibitemOpen
  \bibfield  {author} {\bibinfo {author} {\bibfnamefont {R.}~\bibnamefont
  {Ernstorfer}}, \bibinfo {author} {\bibfnamefont {M.}~\bibnamefont {Harb}},
  \bibinfo {author} {\bibfnamefont {C.~T.}\ \bibnamefont {Hebeisen}}, \bibinfo
  {author} {\bibfnamefont {G.}~\bibnamefont {Sciaini}}, \bibinfo {author}
  {\bibfnamefont {T.}~\bibnamefont {Dartigalongue}}, \ and\ \bibinfo {author}
  {\bibfnamefont {R.~J.~D.}\ \bibnamefont {Miller}},\ }\href {\doibase
  10.1126/science.1162697} {\bibfield  {journal} {\bibinfo  {journal}
  {Science}\ }\textbf {\bibinfo {volume} {323}},\ \bibinfo {pages} {1033}
  (\bibinfo {year} {2009})}\BibitemShut {NoStop}%
\bibitem [{\citenamefont {Eichberger}\ \emph {et~al.}(2010)\citenamefont
  {Eichberger}, \citenamefont {Sch\"{a}fer}, \citenamefont {Krumova},
  \citenamefont {Beyer}, \citenamefont {Demsar}, \citenamefont {Berger},
  \citenamefont {Moriena}, \citenamefont {Sciaini},\ and\ \citenamefont
  {Miller}}]{Eichberger2010}%
  \BibitemOpen
  \bibfield  {author} {\bibinfo {author} {\bibfnamefont {M.}~\bibnamefont
  {Eichberger}}, \bibinfo {author} {\bibfnamefont {H.}~\bibnamefont
  {Sch\"{a}fer}}, \bibinfo {author} {\bibfnamefont {M.}~\bibnamefont
  {Krumova}}, \bibinfo {author} {\bibfnamefont {M.}~\bibnamefont {Beyer}},
  \bibinfo {author} {\bibfnamefont {J.}~\bibnamefont {Demsar}}, \bibinfo
  {author} {\bibfnamefont {H.}~\bibnamefont {Berger}}, \bibinfo {author}
  {\bibfnamefont {G.}~\bibnamefont {Moriena}}, \bibinfo {author} {\bibfnamefont
  {G.}~\bibnamefont {Sciaini}}, \ and\ \bibinfo {author} {\bibfnamefont
  {R.~J.~D.}\ \bibnamefont {Miller}},\ }\href {\doibase 10.1038/nature09539}
  {\bibfield  {journal} {\bibinfo  {journal} {Nature}\ }\textbf {\bibinfo
  {volume} {468}},\ \bibinfo {pages} {799} (\bibinfo {year}
  {2010})}\BibitemShut {NoStop}%
\bibitem [{\citenamefont {Morrison}\ \emph {et~al.}(2014)\citenamefont
  {Morrison}, \citenamefont {Chatelain}, \citenamefont {Tiwari}, \citenamefont
  {Hendaoui}, \citenamefont {Bruhacs}, \citenamefont {Chaker},\ and\
  \citenamefont {Siwick}}]{Morrison2014}%
  \BibitemOpen
  \bibfield  {author} {\bibinfo {author} {\bibfnamefont {V.~R.}\ \bibnamefont
  {Morrison}}, \bibinfo {author} {\bibfnamefont {R.~P.}\ \bibnamefont
  {Chatelain}}, \bibinfo {author} {\bibfnamefont {K.~L.}\ \bibnamefont
  {Tiwari}}, \bibinfo {author} {\bibfnamefont {A.}~\bibnamefont {Hendaoui}},
  \bibinfo {author} {\bibfnamefont {A.}~\bibnamefont {Bruhacs}}, \bibinfo
  {author} {\bibfnamefont {M.}~\bibnamefont {Chaker}}, \ and\ \bibinfo {author}
  {\bibfnamefont {B.~J.}\ \bibnamefont {Siwick}},\ }\href {\doibase
  10.1126/science.1253779} {\bibfield  {journal} {\bibinfo  {journal}
  {Science}\ }\textbf {\bibinfo {volume} {346}},\ \bibinfo {pages} {445}
  (\bibinfo {year} {2014})}\BibitemShut {NoStop}%
\bibitem [{\citenamefont {Park}\ \emph {et~al.}(2005)\citenamefont {Park},
  \citenamefont {Wang}, \citenamefont {Nie}, \citenamefont {Clinite},\ and\
  \citenamefont {Cao}}]{Park2005}%
  \BibitemOpen
  \bibfield  {author} {\bibinfo {author} {\bibfnamefont {H.}~\bibnamefont
  {Park}}, \bibinfo {author} {\bibfnamefont {X.}~\bibnamefont {Wang}}, \bibinfo
  {author} {\bibfnamefont {S.}~\bibnamefont {Nie}}, \bibinfo {author}
  {\bibfnamefont {R.}~\bibnamefont {Clinite}}, \ and\ \bibinfo {author}
  {\bibfnamefont {J.}~\bibnamefont {Cao}},\ }\href {\doibase
  10.1103/PhysRevB.72.100301} {\bibfield  {journal} {\bibinfo  {journal}
  {Physical Review B}\ }\textbf {\bibinfo {volume} {72}},\ \bibinfo {pages}
  {100301} (\bibinfo {year} {2005})}\BibitemShut {NoStop}%
\bibitem [{\citenamefont {Williamson}, \citenamefont {Mourou},\ and\
  \citenamefont {Li}(1984)}]{Williamson1984}%
  \BibitemOpen
  \bibfield  {author} {\bibinfo {author} {\bibfnamefont {S.}~\bibnamefont
  {Williamson}}, \bibinfo {author} {\bibfnamefont {G.}~\bibnamefont {Mourou}},
  \ and\ \bibinfo {author} {\bibfnamefont {J.}~\bibnamefont {Li}},\ }\href
  {https://journals.aps.org/prl/pdf/10.1103/PhysRevLett.52.2364} {\bibfield
  {journal} {\bibinfo  {journal} {Physical Review Letters}\ }\textbf {\bibinfo
  {volume} {52}} (\bibinfo {year} {1984})}\BibitemShut {NoStop}%
\bibitem [{\citenamefont {Dantus}\ \emph {et~al.}(1994)\citenamefont {Dantus},
  \citenamefont {Kim}, \citenamefont {Williamson},\ and\ \citenamefont
  {Zewail}}]{Dantus1994}%
  \BibitemOpen
  \bibfield  {author} {\bibinfo {author} {\bibfnamefont {M.}~\bibnamefont
  {Dantus}}, \bibinfo {author} {\bibfnamefont {S.~B.}\ \bibnamefont {Kim}},
  \bibinfo {author} {\bibfnamefont {J.~C.}\ \bibnamefont {Williamson}}, \ and\
  \bibinfo {author} {\bibfnamefont {A.~H.}\ \bibnamefont {Zewail}},\ }\href
  {\doibase 10.1021/j100062a011} {\bibfield  {journal} {\bibinfo  {journal}
  {Journal of Physical Chemistry}\ }\textbf {\bibinfo {volume} {98}},\ \bibinfo
  {pages} {2782} (\bibinfo {year} {1994})}\BibitemShut {NoStop}%
\bibitem [{\citenamefont {Kassier}\ \emph {et~al.}(2010)\citenamefont
  {Kassier}, \citenamefont {Haupt}, \citenamefont {Erasmus}, \citenamefont
  {Rohwer}, \citenamefont {von Bergmann}, \citenamefont {Schwoerer},
  \citenamefont {Coelho},\ and\ \citenamefont {Auret}}]{Kassier2010}%
  \BibitemOpen
  \bibfield  {author} {\bibinfo {author} {\bibfnamefont {G.~H.}\ \bibnamefont
  {Kassier}}, \bibinfo {author} {\bibfnamefont {K.}~\bibnamefont {Haupt}},
  \bibinfo {author} {\bibfnamefont {N.}~\bibnamefont {Erasmus}}, \bibinfo
  {author} {\bibfnamefont {E.~G.}\ \bibnamefont {Rohwer}}, \bibinfo {author}
  {\bibfnamefont {H.~M.}\ \bibnamefont {von Bergmann}}, \bibinfo {author}
  {\bibfnamefont {H.}~\bibnamefont {Schwoerer}}, \bibinfo {author}
  {\bibfnamefont {S.~M.~M.}\ \bibnamefont {Coelho}}, \ and\ \bibinfo {author}
  {\bibfnamefont {F.~D.}\ \bibnamefont {Auret}},\ }\href {\doibase
  10.1063/1.3489118} {\bibfield  {journal} {\bibinfo  {journal} {Review of
  Scientific Instruments}\ }\textbf {\bibinfo {volume} {81}},\ \bibinfo {pages}
  {105103} (\bibinfo {year} {2010})}\BibitemShut {NoStop}%
\bibitem [{\citenamefont {Hebeisen}\ \emph {et~al.}(2008)\citenamefont
  {Hebeisen}, \citenamefont {Sciaini}, \citenamefont {Harb}, \citenamefont
  {Ernstorfer}, \citenamefont {Dartigalongue}, \citenamefont {Kruglik},\ and\
  \citenamefont {Miller}}]{Hebeisen2008}%
  \BibitemOpen
  \bibfield  {author} {\bibinfo {author} {\bibfnamefont {C.~T.}\ \bibnamefont
  {Hebeisen}}, \bibinfo {author} {\bibfnamefont {G.}~\bibnamefont {Sciaini}},
  \bibinfo {author} {\bibfnamefont {M.}~\bibnamefont {Harb}}, \bibinfo {author}
  {\bibfnamefont {R.}~\bibnamefont {Ernstorfer}}, \bibinfo {author}
  {\bibfnamefont {T.}~\bibnamefont {Dartigalongue}}, \bibinfo {author}
  {\bibfnamefont {S.~G.}\ \bibnamefont {Kruglik}}, \ and\ \bibinfo {author}
  {\bibfnamefont {R.~J.~D.}\ \bibnamefont {Miller}},\ }\href
  {http://www.ncbi.nlm.nih.gov/pubmed/18542423} {\bibfield  {journal} {\bibinfo
   {journal} {Optics Express}\ }\textbf {\bibinfo {volume} {16}},\ \bibinfo
  {pages} {3334} (\bibinfo {year} {2008})}\BibitemShut {NoStop}%
\bibitem [{\citenamefont {Aidelsburger}\ \emph {et~al.}(2010)\citenamefont
  {Aidelsburger}, \citenamefont {Kirchner}, \citenamefont {Krausz},\ and\
  \citenamefont {Baum}}]{Aidelsburger2010}%
  \BibitemOpen
  \bibfield  {author} {\bibinfo {author} {\bibfnamefont {M.}~\bibnamefont
  {Aidelsburger}}, \bibinfo {author} {\bibfnamefont {F.~O.}\ \bibnamefont
  {Kirchner}}, \bibinfo {author} {\bibfnamefont {F.}~\bibnamefont {Krausz}}, \
  and\ \bibinfo {author} {\bibfnamefont {P.}~\bibnamefont {Baum}},\ }\href
  {\doibase 10.1073/pnas.1010165107} {\bibfield  {journal} {\bibinfo  {journal}
  {PNAS}\ }\textbf {\bibinfo {volume} {107}},\ \bibinfo {pages} {19714}
  (\bibinfo {year} {2010})}\BibitemShut {NoStop}%
\bibitem [{\citenamefont {Siwick}\ \emph {et~al.}(2002)\citenamefont {Siwick},
  \citenamefont {Dwyer}, \citenamefont {Jordan},\ and\ \citenamefont
  {Miller}}]{Siwick2002}%
  \BibitemOpen
  \bibfield  {author} {\bibinfo {author} {\bibfnamefont {B.~J.}\ \bibnamefont
  {Siwick}}, \bibinfo {author} {\bibfnamefont {J.~R.}\ \bibnamefont {Dwyer}},
  \bibinfo {author} {\bibfnamefont {R.~E.}\ \bibnamefont {Jordan}}, \ and\
  \bibinfo {author} {\bibfnamefont {R.~J.~D.}\ \bibnamefont {Miller}},\ }\href
  {\doibase 10.1063/1.1487437} {\bibfield  {journal} {\bibinfo  {journal}
  {Journal of Applied Physics}\ }\textbf {\bibinfo {volume} {92}},\ \bibinfo
  {pages} {1643} (\bibinfo {year} {2002})}\BibitemShut {NoStop}%
\bibitem [{\citenamefont {Fill}\ \emph {et~al.}(2006)\citenamefont {Fill},
  \citenamefont {Veisz}, \citenamefont {Apolonski},\ and\ \citenamefont
  {Krausz}}]{Fill2006}%
  \BibitemOpen
  \bibfield  {author} {\bibinfo {author} {\bibfnamefont {E.}~\bibnamefont
  {Fill}}, \bibinfo {author} {\bibfnamefont {L.}~\bibnamefont {Veisz}},
  \bibinfo {author} {\bibfnamefont {A.}~\bibnamefont {Apolonski}}, \ and\
  \bibinfo {author} {\bibfnamefont {F.}~\bibnamefont {Krausz}},\ }\href
  {\doibase 10.1088/1367-2630/8/11/272} {\bibfield  {journal} {\bibinfo
  {journal} {New Journal of Physics}\ }\textbf {\bibinfo {volume} {8}},\
  \bibinfo {pages} {272} (\bibinfo {year} {2006})}\BibitemShut {NoStop}%
\bibitem [{\citenamefont {van Oudheusden}\ \emph {et~al.}(2007)\citenamefont
  {van Oudheusden}, \citenamefont {de~Jong}, \citenamefont {van~der Geer},
  \citenamefont {{'t Root}}, \citenamefont {Luiten},\ and\ \citenamefont
  {Siwick}}]{VanOudheusden2007}%
  \BibitemOpen
  \bibfield  {author} {\bibinfo {author} {\bibfnamefont {T.}~\bibnamefont {van
  Oudheusden}}, \bibinfo {author} {\bibfnamefont {E.~F.}\ \bibnamefont
  {de~Jong}}, \bibinfo {author} {\bibfnamefont {S.~B.}\ \bibnamefont {van~der
  Geer}}, \bibinfo {author} {\bibfnamefont {W.~P. E. M.~O.}\ \bibnamefont {{'t
  Root}}}, \bibinfo {author} {\bibfnamefont {O.~J.}\ \bibnamefont {Luiten}}, \
  and\ \bibinfo {author} {\bibfnamefont {B.~J.}\ \bibnamefont {Siwick}},\
  }\href {\doibase 10.1063/1.2801027} {\bibfield  {journal} {\bibinfo
  {journal} {Journal of Applied Physics}\ }\textbf {\bibinfo {volume} {102}},\
  \bibinfo {pages} {093501} (\bibinfo {year} {2007})}\BibitemShut {NoStop}%
\bibitem [{\citenamefont {Kassier}\ \emph {et~al.}(2009)\citenamefont
  {Kassier}, \citenamefont {Haupt}, \citenamefont {Erasmus}, \citenamefont
  {Rohwer},\ and\ \citenamefont {Schwoerer}}]{Kassier2009}%
  \BibitemOpen
  \bibfield  {author} {\bibinfo {author} {\bibfnamefont {G.~H.}\ \bibnamefont
  {Kassier}}, \bibinfo {author} {\bibfnamefont {K.}~\bibnamefont {Haupt}},
  \bibinfo {author} {\bibfnamefont {N.}~\bibnamefont {Erasmus}}, \bibinfo
  {author} {\bibfnamefont {E.~G.}\ \bibnamefont {Rohwer}}, \ and\ \bibinfo
  {author} {\bibfnamefont {H.}~\bibnamefont {Schwoerer}},\ }\href {\doibase
  10.1063/1.3132834} {\bibfield  {journal} {\bibinfo  {journal} {Journal of
  Applied Physics}\ }\textbf {\bibinfo {volume} {105}},\ \bibinfo {pages}
  {113111} (\bibinfo {year} {2009})}\BibitemShut {NoStop}%
\bibitem [{\citenamefont {Chatelain}\ \emph {et~al.}(2012)\citenamefont
  {Chatelain}, \citenamefont {Morrison}, \citenamefont {Godbout},\ and\
  \citenamefont {Siwick}}]{Chatelain2012}%
  \BibitemOpen
  \bibfield  {author} {\bibinfo {author} {\bibfnamefont {R.~P.}\ \bibnamefont
  {Chatelain}}, \bibinfo {author} {\bibfnamefont {V.~R.}\ \bibnamefont
  {Morrison}}, \bibinfo {author} {\bibfnamefont {C.}~\bibnamefont {Godbout}}, \
  and\ \bibinfo {author} {\bibfnamefont {B.~J.}\ \bibnamefont {Siwick}},\
  }\href {\doibase 10.1063/1.4747155} {\bibfield  {journal} {\bibinfo
  {journal} {Applied Physics Letters}\ }\textbf {\bibinfo {volume} {101}},\
  \bibinfo {pages} {081901} (\bibinfo {year} {2012})}\BibitemShut {NoStop}%
\bibitem [{\citenamefont {Gao}\ \emph {et~al.}(2012)\citenamefont {Gao},
  \citenamefont {Jean-Ruel}, \citenamefont {Cooney}, \citenamefont {Stampe},
  \citenamefont {Jong}, \citenamefont {Harb}, \citenamefont {Sciaini},
  \citenamefont {Moriena},\ and\ \citenamefont {Miller}}]{Gao2012}%
  \BibitemOpen
  \bibfield  {author} {\bibinfo {author} {\bibfnamefont {M.}~\bibnamefont
  {Gao}}, \bibinfo {author} {\bibfnamefont {H.}~\bibnamefont {Jean-Ruel}},
  \bibinfo {author} {\bibfnamefont {R.~R.}\ \bibnamefont {Cooney}}, \bibinfo
  {author} {\bibfnamefont {J.}~\bibnamefont {Stampe}}, \bibinfo {author}
  {\bibfnamefont {M.~D.}\ \bibnamefont {Jong}}, \bibinfo {author}
  {\bibfnamefont {M.}~\bibnamefont {Harb}}, \bibinfo {author} {\bibfnamefont
  {G.}~\bibnamefont {Sciaini}}, \bibinfo {author} {\bibfnamefont
  {G.}~\bibnamefont {Moriena}}, \ and\ \bibinfo {author} {\bibfnamefont
  {R.~J.~D.}\ \bibnamefont {Miller}},\ }\href {\doibase 10.1364/OE.20.012048}
  {\bibfield  {journal} {\bibinfo  {journal} {Optics Express}\ }\textbf
  {\bibinfo {volume} {20}},\ \bibinfo {pages} {12048} (\bibinfo {year}
  {2012})}\BibitemShut {NoStop}%
\bibitem [{\citenamefont {Mancini}\ \emph {et~al.}(2012)\citenamefont
  {Mancini}, \citenamefont {Mansart}, \citenamefont {Pagano}, \citenamefont
  {van~der Geer}, \citenamefont {de~Loos},\ and\ \citenamefont
  {Carbone}}]{Mancini2012}%
  \BibitemOpen
  \bibfield  {author} {\bibinfo {author} {\bibfnamefont {G.~F.}\ \bibnamefont
  {Mancini}}, \bibinfo {author} {\bibfnamefont {B.}~\bibnamefont {Mansart}},
  \bibinfo {author} {\bibfnamefont {S.}~\bibnamefont {Pagano}}, \bibinfo
  {author} {\bibfnamefont {B.}~\bibnamefont {van~der Geer}}, \bibinfo {author}
  {\bibfnamefont {M.}~\bibnamefont {de~Loos}}, \ and\ \bibinfo {author}
  {\bibfnamefont {F.}~\bibnamefont {Carbone}},\ }\href {\doibase
  10.1016/j.nima.2012.06.057} {\bibfield  {journal} {\bibinfo  {journal} {Nucl.
  Instrum. Meth. A}\ }\textbf {\bibinfo {volume} {691}},\ \bibinfo {pages}
  {113} (\bibinfo {year} {2012})}\BibitemShut {NoStop}%
\bibitem [{\citenamefont {Wilhelm}, \citenamefont {Piel},\ and\ \citenamefont
  {Riedle}(1997)}]{Wilhelm1997}%
  \BibitemOpen
  \bibfield  {author} {\bibinfo {author} {\bibfnamefont {T.}~\bibnamefont
  {Wilhelm}}, \bibinfo {author} {\bibfnamefont {J.}~\bibnamefont {Piel}}, \
  and\ \bibinfo {author} {\bibfnamefont {E.}~\bibnamefont {Riedle}},\ }\href
  {\doibase 10.1364/OL.22.001494} {\bibfield  {journal} {\bibinfo  {journal}
  {Opt. Lett.}\ }\textbf {\bibinfo {volume} {22}},\ \bibinfo {pages} {1494}
  (\bibinfo {year} {1997})}\BibitemShut {NoStop}%
\bibitem [{\citenamefont {Latham}(1995)}]{book:Latham}%
  \BibitemOpen
  \bibinfo {editor} {\bibfnamefont {R.}~\bibnamefont {Latham}},\ ed.,\
  \href@noop {} {\emph {\bibinfo {title} {High Voltage Vacuum Insulation}}}\
  (\bibinfo  {publisher} {Academic Press},\ \bibinfo {year} {1995})\BibitemShut
  {NoStop}%
\bibitem [{\citenamefont {Bruce}(1947)}]{Bruce1946}%
  \BibitemOpen
  \bibfield  {author} {\bibinfo {author} {\bibfnamefont {F.~M.}\ \bibnamefont
  {Bruce}},\ }\href {\doibase 10.1049/ji-2.1947.0052} {\bibfield  {journal}
  {\bibinfo  {journal} {J. Inst. Electr. Eng., Part 2}\ }\textbf {\bibinfo
  {volume} {94}},\ \bibinfo {pages} {138 } (\bibinfo {year}
  {1947})}\BibitemShut {NoStop}%
\bibitem [{\citenamefont {Michalik}, \citenamefont {Sherman},\ and\
  \citenamefont {Sipe}(2008)}]{Michalik2008}%
  \BibitemOpen
  \bibfield  {author} {\bibinfo {author} {\bibfnamefont {A.~M.}\ \bibnamefont
  {Michalik}}, \bibinfo {author} {\bibfnamefont {E.~Y.}\ \bibnamefont
  {Sherman}}, \ and\ \bibinfo {author} {\bibfnamefont {J.~E.}\ \bibnamefont
  {Sipe}},\ }\href {\doibase 10.1063/1.2973157} {\bibfield  {journal} {\bibinfo
   {journal} {Journal of Applied Physics}\ }\textbf {\bibinfo {volume} {104}},\
  \bibinfo {pages} {054905} (\bibinfo {year} {2008})}\BibitemShut {NoStop}%
\bibitem [{\citenamefont {Waldecker}\ \emph {et~al.}(2014)\citenamefont
  {Waldecker}, \citenamefont {Miller}, \citenamefont {Rude}, \citenamefont
  {Bertoni}, \citenamefont {Osmond}, \citenamefont {Pruneri}, \citenamefont
  {Simpson}, \citenamefont {Ernstorfer},\ and\ \citenamefont
  {Wall}}]{Waldecker2014}%
  \BibitemOpen
  \bibfield  {author} {\bibinfo {author} {\bibfnamefont {L.}~\bibnamefont
  {Waldecker}}, \bibinfo {author} {\bibfnamefont {T.~A.}\ \bibnamefont
  {Miller}}, \bibinfo {author} {\bibfnamefont {M.}~\bibnamefont {Rude}},
  \bibinfo {author} {\bibfnamefont {R.}~\bibnamefont {Bertoni}}, \bibinfo
  {author} {\bibfnamefont {J.}~\bibnamefont {Osmond}}, \bibinfo {author}
  {\bibfnamefont {V.}~\bibnamefont {Pruneri}}, \bibinfo {author} {\bibfnamefont
  {R.}~\bibnamefont {Simpson}}, \bibinfo {author} {\bibfnamefont
  {R.}~\bibnamefont {Ernstorfer}}, \ and\ \bibinfo {author} {\bibfnamefont
  {S.}~\bibnamefont {Wall}},\ }\href {http://arxiv.org/abs/1412.0901}
  {\bibfield  {journal} {\bibinfo  {journal} {ArXiv preprint arXiv:1412.0901}\
  } (\bibinfo {year} {2014})}\BibitemShut {NoStop}%
\bibitem [{\citenamefont {Dwyer}\ \emph {et~al.}(2006)\citenamefont {Dwyer},
  \citenamefont {Hebeisen}, \citenamefont {Ernstorfer}, \citenamefont {Harb},
  \citenamefont {Deyirmenjian}, \citenamefont {Jordan},\ and\ \citenamefont
  {Miller}}]{Dwyer2006}%
  \BibitemOpen
  \bibfield  {author} {\bibinfo {author} {\bibfnamefont {J.~R.}\ \bibnamefont
  {Dwyer}}, \bibinfo {author} {\bibfnamefont {C.~T.}\ \bibnamefont {Hebeisen}},
  \bibinfo {author} {\bibfnamefont {R.}~\bibnamefont {Ernstorfer}}, \bibinfo
  {author} {\bibfnamefont {M.}~\bibnamefont {Harb}}, \bibinfo {author}
  {\bibfnamefont {V.~B.}\ \bibnamefont {Deyirmenjian}}, \bibinfo {author}
  {\bibfnamefont {R.~E.}\ \bibnamefont {Jordan}}, \ and\ \bibinfo {author}
  {\bibfnamefont {R.~J.~D.}\ \bibnamefont {Miller}},\ }\href {\doibase
  10.1098/rsta.2005.1735} {\bibfield  {journal} {\bibinfo  {journal} {Philos
  Trans A Math Phys Eng Sci}\ }\textbf {\bibinfo {volume} {364}},\ \bibinfo
  {pages} {741} (\bibinfo {year} {2006})}\BibitemShut {NoStop}%
\bibitem [{\citenamefont {Gerbig}\ \emph {et~al.}(2012)\citenamefont {Gerbig},
  \citenamefont {Morgenstern}, \citenamefont {Sarpe}, \citenamefont
  {Wollenhaupt},\ and\ \citenamefont {Baumert}}]{Gerbig2012}%
  \BibitemOpen
  \bibfield  {author} {\bibinfo {author} {\bibfnamefont {C.}~\bibnamefont
  {Gerbig}}, \bibinfo {author} {\bibfnamefont {S.}~\bibnamefont {Morgenstern}},
  \bibinfo {author} {\bibfnamefont {C.}~\bibnamefont {Sarpe}}, \bibinfo
  {author} {\bibfnamefont {M.}~\bibnamefont {Wollenhaupt}}, \ and\ \bibinfo
  {author} {\bibfnamefont {T.}~\bibnamefont {Baumert}},\ }in\ \href {\doibase
  10.1364/ICUSD.2012.IT3D.3} {\emph {\bibinfo {booktitle} {Research in Optical
  Sciences}}}\ (\bibinfo  {publisher} {Osa},\ \bibinfo {address} {Washington,
  D.C.},\ \bibinfo {year} {2012})\ p.\ \bibinfo {pages} {IT3D.3}\BibitemShut
  {NoStop}%
\bibitem [{\citenamefont {Weninger}\ and\ \citenamefont
  {Baum}(2012)}]{Weninger2012}%
  \BibitemOpen
  \bibfield  {author} {\bibinfo {author} {\bibfnamefont {C.}~\bibnamefont
  {Weninger}}\ and\ \bibinfo {author} {\bibfnamefont {P.}~\bibnamefont
  {Baum}},\ }\href {\doibase 10.1016/j.ultramic.2011.11.018} {\bibfield
  {journal} {\bibinfo  {journal} {Ultramicroscopy}\ }\textbf {\bibinfo {volume}
  {113}},\ \bibinfo {pages} {145} (\bibinfo {year} {2012})}\BibitemShut
  {NoStop}%
\bibitem [{\citenamefont {Kirchner}\ \emph {et~al.}(2013)\citenamefont
  {Kirchner}, \citenamefont {Lahme}, \citenamefont {Krausz},\ and\
  \citenamefont {Baum}}]{Kirchner2013}%
  \BibitemOpen
  \bibfield  {author} {\bibinfo {author} {\bibfnamefont {F.~O.}\ \bibnamefont
  {Kirchner}}, \bibinfo {author} {\bibfnamefont {S.}~\bibnamefont {Lahme}},
  \bibinfo {author} {\bibfnamefont {F.}~\bibnamefont {Krausz}}, \ and\ \bibinfo
  {author} {\bibfnamefont {P.}~\bibnamefont {Baum}},\ }\href {\doibase
  10.1088/1367-2630/15/6/063021} {\bibfield  {journal} {\bibinfo  {journal}
  {New Journal of Physics}\ }\textbf {\bibinfo {volume} {15}},\ \bibinfo
  {pages} {063021} (\bibinfo {year} {2013})}\BibitemShut {NoStop}%
\bibitem [{\citenamefont {Qian}\ and\ \citenamefont
  {Elsayed-Ali}(2003)}]{Qian2003}%
  \BibitemOpen
  \bibfield  {author} {\bibinfo {author} {\bibfnamefont {B.-L.}\ \bibnamefont
  {Qian}}\ and\ \bibinfo {author} {\bibfnamefont {H.~E.}\ \bibnamefont
  {Elsayed-Ali}},\ }\href {\doibase 10.1063/1.1567816} {\bibfield  {journal}
  {\bibinfo  {journal} {Journal of Applied Physics}\ }\textbf {\bibinfo
  {volume} {94}},\ \bibinfo {pages} {803} (\bibinfo {year} {2003})}\BibitemShut
  {NoStop}%
\end{thebibliography}%

\end{document}